\documentclass[aps,prb,onecolumn,floatfix,superscriptaddress,11pt]{revtex4}

\usepackage[normalem]{ulem}

\usepackage{fancyhdr}
\usepackage{kantlipsum}

\usepackage{upgreek}

\usepackage{amsmath}
\usepackage{graphicx}

\usepackage{color}

\usepackage{floatrow}

\usepackage[unicode=true,bookmarks=false,breaklinks=false,pdfborder={0  0 1},backref=false,colorlinks=false]{hyperref} 
\usepackage{breakurl}

\usepackage{lettrine}

\setcitestyle{round}

\makeatletter

\newcommand{\nanotec}{
 CNR NANOTEC − Institute of Nanotechnology, via Monteroni, 73100, Lecce, Italy
}
\newcommand{\unisal}{
 Dipartimento di matematica e fisica ''Ennio De Giorgi'', Universit\`{a} del Salento, Via Arnesano, 73100 Lecce, Italy
}
\newcommand{\canada}{
 Department of Engineering Physics, \'{E}cole Polytechnique de Montr\'{e}al, Montr\'{e}al, Qu\'{e}bec H3C 3A7, Canada
}
\newcommand{\finland}{
 COMP Centre of Excellence, Department of Applied Physics, Aalto University, PO Box 15100, Fi-00076 Aalto, Finland
}
\newcommand{\ucl}{
 Department of Physics, Imperial College London, London SW7 2AZ, United Kingdom
}

\begin{document}

\title{
{
{\Large
\textbf{
Room-temperature superfluidity in a polariton condensate
}
}
}
}

\author{Giovanni Lerario}
\affiliation{\nanotec}

\author{Antonio Fieramosca}
\affiliation{\nanotec}
\affiliation{\unisal}

\author{F\'{a}bio Barachati}
\affiliation{\canada}

\author{Dario Ballarini}
\email{dario.ballarini@gmail.com}
\affiliation{\nanotec}

\author{Konstantinos S. Daskalakis}
\affiliation{\finland}

\author{Lorenzo Dominici}
\affiliation{\nanotec}

\author{Milena De Giorgi}
\affiliation{\nanotec}

\author{Stefan A. Maier}
\affiliation{\ucl}

\author{Giuseppe Gigli}
\affiliation{\nanotec}
\affiliation{\unisal}

\author{St\'{e}phane K\'{e}na-Cohen}
\affiliation{\canada}

\author{Daniele Sanvitto}
\affiliation{\nanotec}

\begin{abstract}

\end{abstract}

\maketitle

\noindent\textbf{Superfluidity---the suppression of scattering in a quantum fluid at velocities below a critical value---is one of the most striking manifestations of the collective behaviour typical of Bose-Einstein condensates. This phenomenon, akin to superconductivity in metals, has until now only been observed at prohibitively low cryogenic temperatures. For atoms, this limit is imposed by the small thermal de Broglie wavelength, which is inversely related to the particle mass. Even in the case of ultralight quasiparticles such as exciton-polaritons, superfluidity has only been demonstrated at liquid helium temperatures. In this case, the limit is not imposed by the mass, but instead by the small exciton binding energy of Wannier-Mott excitons, which places the upper temperature limit. Here we demonstrate a transition from normal to superfluid flow in an organic microcavity supporting stable Frenkel exciton-polaritons at room temperature. This result paves the way not only to table-top studies of quantum hydrodynamics, but also to room-temperature polariton devices that can be robustly protected from scattering. 
}

\vspace{0.3cm}

\noindent{\textbf{Keywords:} superfluidity, microcavities, polaritons, BEC, condensate\\
}

\noindent 
First observed in liquid helium below the lambda point, superfluidity manifests itself in a number of fascinating ways. In the superfluid phase, helium can creep up along the walls of a container, boil without bubbles or even flow without friction around obstacles. As early as 1938, Fritz London suggested a link between superfluidity and Bose-Einstein condensation (BEC)~\cite{london_-phenomenon_1938}. Indeed, superfluidity is now known to be related to the finite amount of energy needed to create collective excitations in the quantum liquid~\cite{allen_flow_1938,kapitza_viscosity_1938,landau_theory_1941,landau_theory_1949} and the link proposed by London was further evidenced by the observation of superfluidity in ultracold atomic BECs~\cite{onofrio_observation_2000,desbuquois_superfluid_2012}. A quantitative description is given by the Gross-Pitaevskii (GP) equation~\cite{gross_structure_1961,pitaevskii_vortex_1961} (see Methods) and the perturbation theory for elementary excitations developed by Bogoliubov~\cite{pitaevskii_bose-einstein_2003}. First derived for atomic condensates this theory has since been successfully applied to a variety of systems and the mathematical framework of the GP equation naturally leads to important analogies between BEC and nonlinear optics~\cite{bolda_dissipative_2001,carusotto_quantum_2013}. Recently, it has been extended to include condensates far from thermal equilibrium composed of bosonic quasi-particles such as microcavity exciton-polaritons, magnons and interacting photons\cite{carusotto_quantum_2013,carusotto_probing_2004}. In particular, for exciton-polaritons, the observation of many-body effects related to condensation and superfluidity, such as the excitation of quantized vortices, the formation of metastable currents and the suppression of scattering from potential barriers~\cite{lagoudakis_quantized_2008,sanvitto_persistent_2010,nardin_hydrodynamic_2011,sanvitto_all-optical_2011,amo_polariton_2011,amo_superfluidity_2009}, have shown the rich phenomenology that can also exist within non-equilibrium condensates. However, until now, all these phenomena have been mainly observed in microcavities embedding quantum wells of III-V or II-VI semiconductors. As a result, experiments must be performed at low temperatures (below $\sim$20 K), beyond which excitons autoionize. This is a consequence of the low binding energy typical of Wannier-Mott excitons. Frenkel excitons, which are characteristic of organic semiconductors, possess large binding energies that readily allow for strong light-matter coupling and the formation of polaritons at room temperature. Remarkably, in spite of the weaker interactions between organic polaritons as compared to inorganic ones, condensation and the spontaneous formation of vortices have recently been observed in these materials~\cite{plumhof_room-temperature_2013,daskalakis_nonlinear_2014,daskalakis_spatial_2015}.  However, the small polariton-polariton interaction constants, structural inhomogeneity and short lifetimes in these structures have until now prevented the observation of behaviour directly related to the quantum fluid dynamics such as superfluidity. In this work, we show that indeed superfluidity can be achieved at room-temperature and this is, in part, a result of the much larger polariton densities achievable in organic microcavities, which compensate for their weaker nonlinearities. These observations pave the way for future table-top studies of quantum hydrodynamics, but also towards eliminating scattering losses in room-temperature polariton circuits.

Our sample consists of an optical microcavity composed of two dielectric mirrors surrounding a thin film of the 2,7-Bis[9,9-di(4-methylphenyl)-fluoren-2-yl]-9,9-di(4-methylphenyl)fluorene (TDAF) organic molecule. Light-matter interaction in this system is so strong that it leads to the formation of hybrid light-matter modes---the polaritons---with a Rabi energy $2\Omega_R \sim 0.6~\text{eV}$. The same structure has been used previously to demonstrate polariton condensation under high energy non-resonant excitation~\cite{daskalakis_nonlinear_2014}. It shows clear nonlinearities, as evidenced by a power-dependent blueshift, and good spatial homogeneity. Under resonant excitation, which consists of choosing the appropriate energy and wavevector from the dispersion relation to directly excite a specific polariton state, it allows for the injection and flow of polaritons with a well-defined group velocity. When such a flow is incident upon a defect, we observe a suppression of elastic scattering in real and momentum space when increasing the density of the lower polariton branch. This suppression is apparent in the strong reduction of density modulations ahead of the defect and in the filling of the dark trail in the wake of the moving fluid.  We find that the transition to a non-viscous fluid manifests itself as a smooth crossover, rather than a sharp threshold-like behaviour~\cite{cancellieri_superflow_2010}. Our results are in good agreement with a time-dependent solution of the GP equation under pulsed excitation. We show that even for the very short polariton lifetime characteristic of our sample, the transition from supersonic to superfluid flow in our calculation occurs at densities comparable to those in our experiments. Indeed, such superfluid propagation in the case of pulsed excitation is very closely related to the original proposal for superfluidity in nonlinear microcavities~\cite{bolda_dissipative_2001}. Although in principle there is no stationary state in the pulsed case, we can nevertheless make the link with the Bogoliubov excitation spectrum of the polariton condensate (see Fig.~S1).  The densities required to achieve superfluidity are approximately those for which the sound speed of elementary excitations is larger than the flow velocity. Given the broadband pump, the polariton density is dominated by the component of the pump which is always resonant with the renormalized dispersion. It is also important to stress that using ultrafast pulsed excitation, we avoid several drawbacks as compared to the resonant steady state used in Ref.~\citenum{amo_superfluidity_2009} and to the perturbation of the parametric process (TOPO) as in Refs.~\citenum{amo_collective_2009, berceanu_multicomponent_2015}. In these earlier works, which were the first demonstration of the superfluid behaviour of polaritons at 4 K, the phase stiffness imposed by the continuous wave pumping laser in the case of resonant excitation~\cite{bolda_dissipative_2001} and the presence of more than one condensed state in the TOPO case~\cite{berceanu_multicomponent_2015} complicated the interpretation of the evolution of the polariton fluid.\\

\noindent
\textbf{Esperimental results\\}
The sample was positioned between two microscope objectives as shown in Fig.~1a to allow for measurements in a transmission geometry while maintaining high spatial resolution. A polariton wavepacket with a chosen wavevector was then created by exciting the sample with a linearly polarized 35 fs laser pulse resonant with the lower polariton branch of the microcavity (see Methods).

\begin{figure}[htbp]
  \centering \includegraphics[width=16cm]{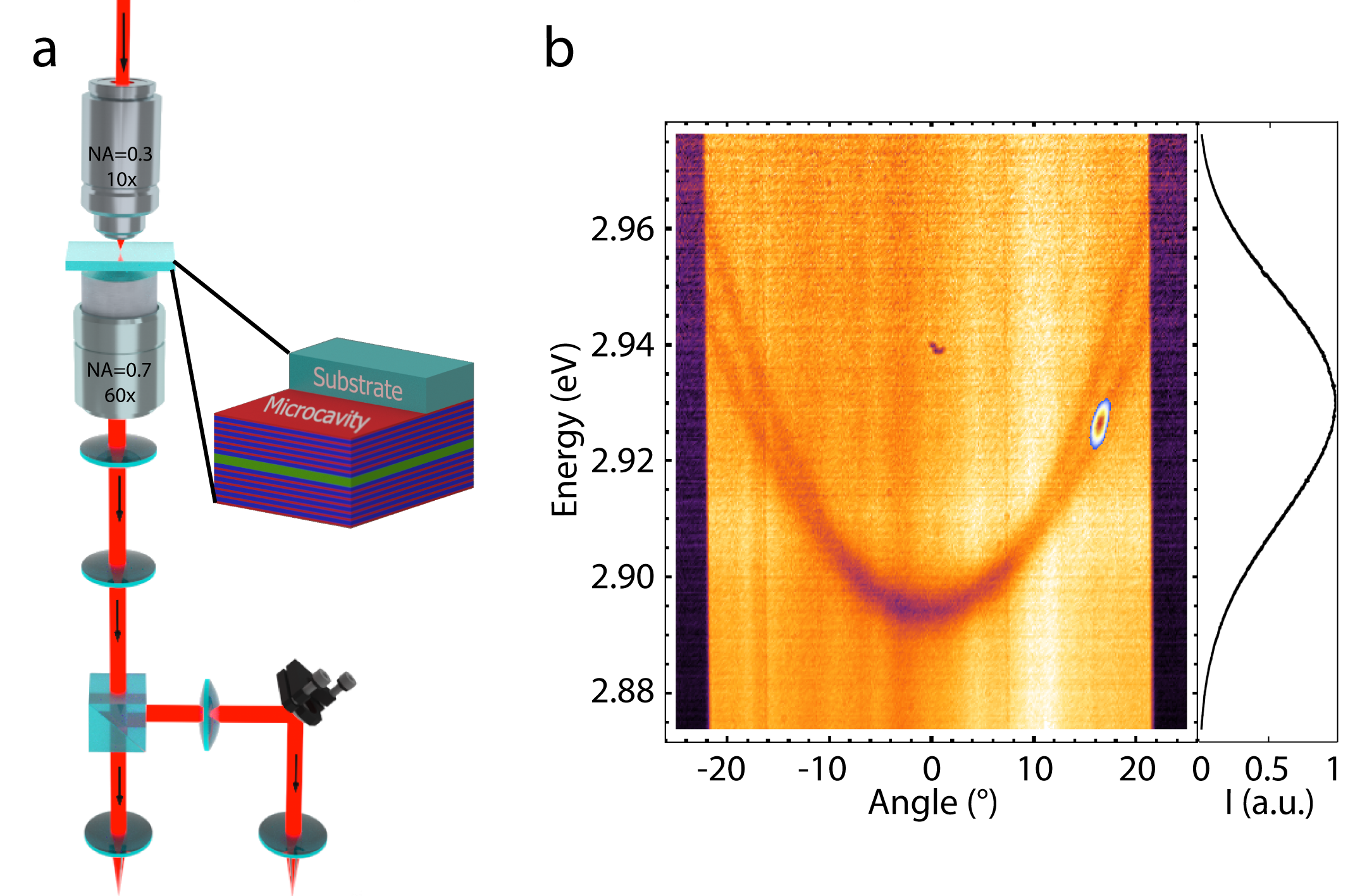} \linespread{1.1}
  \protect\protect\caption{
\textbf{Optical setup and sample dispersion.} \textbf{a,} Sketch of the optical setup. The microcavity is placed between two microscope objectives and the excitation ($35~\text{fs}$ pulsed laser) is incident from the substrate side, while detection is performed from the microcavity deposition side. \textbf{b,} Energy-angle map of the microcavity under white light illumination superimposed with the transmitted signal under resonant laser excitation (the transparency threshold is set at 50\% of the maximum intensity). The incident angle can be related to the in-plane wavevector using $k_{\!/\mkern-5mu/\!} = (\omega/c)\sin{\theta}$.~On the right side, the energy spectrum of the excitation laser is shown. 
  }
\label{fig:FIG1} 
\end{figure}

By detecting the reflected or transmitted light using a spectrometer and CCD camera, energy-resolved space and momentum maps could be acquired. An example of the experimental polariton dispersion under white light illumination is given in Fig.~1b. The parabolic TE- and TM-polarized lower polariton branches appear as dips in the reflectance spectra. The figure also shows an example of how the laser energy, momentum and polarization can be precisely tuned to excite, in this case, the TE lower polariton branch at a given angle. The resonantly excited polariton state is shown as a transmission spot superimposed to the reflectance map. Note the spectral filtering operated by the microcavity itself on the ultrashort pump pulse, shown on the right side of Fig.~1b. In contrast to experiments on superfluidity in atomic condensates~\cite{onofrio_observation_2000,desbuquois_superfluid_2012}, here the position of the defect is kept fixed in space, while the speed of the quasi-particle flow is controlled by using the excitation angle, which is directly related to the polariton group velocity $v_g=\hbar k_p / m_{LP}$, where $k_p$ is the polariton wavevector and $m_{LP}$ the effective mass.

\begin{figure}[htbp]
  \centering \includegraphics[width=11cm]{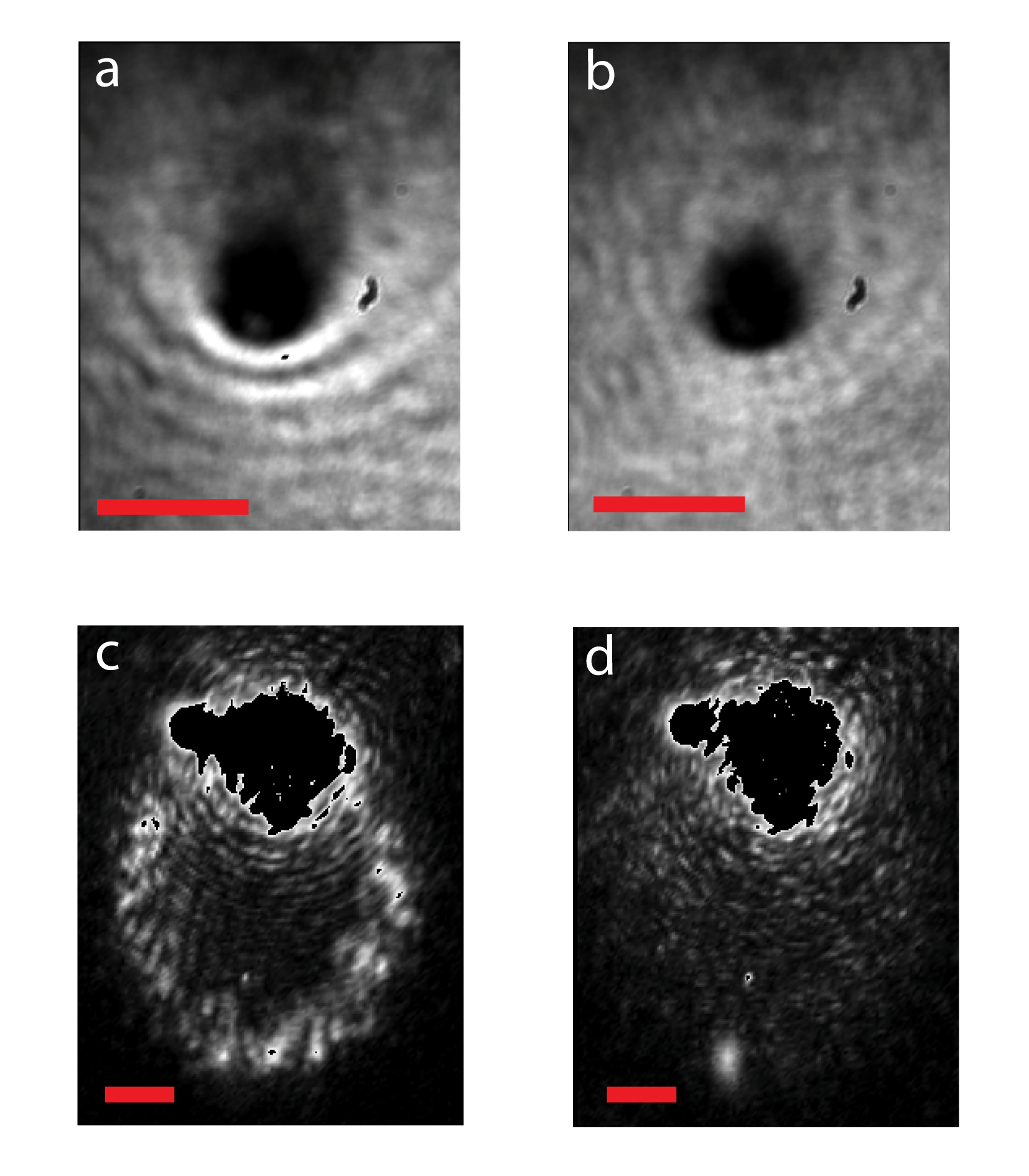} \linespread{1.1}
  \protect\protect\caption{
\textbf{Superfluid behaviour.} \textbf{a,} Real space intensity distribution of the polariton flux with group velocity $19~\upmu \text{m/ps}$ flowing from the bottom to the top of the image across an artificial defect. The polariton density is $0.5 \cdot 10^6~\text{pol}/\upmu \text{m}^2$ and the scale bar is $5~\upmu \text{m}$. The interference pattern is generated by the polaritons scattered at the defect position. A shadow cone beyond the defect reflects the reduction in the number of in-plane transmitted polaritons. \textbf{b,} Real space intensity distribution at polariton densities of $10^7~\text{pol}/\upmu \text{m}^2$ across the same defect. The interference fringes and the shadow cone beyond the defect vanish almost perfectly, revealing the physics of a fluid with zero viscosity. \textbf{c,} Saturated image in momentum space associated to panel a; the elastic scattering ring generated by the presence of the defect is clearly observable. \textbf{d,} Saturated image of the momentum space associated to panel b; the elastic scattering ring is completely suppressed at high polariton densities. The scale bar is $2~\upmu \text{m}^{-1}$.
 }
\label{fig:FIG2} 
\end{figure}

Figure 2 shows a real space image of the time-averaged transmitted (a,b) and scattered (c,d) intensities for a polariton flow moving upwards across an obstacle with an exciting angle chosen such that the flow speed is $19~\upmu \text{m/ps}$. The point defect, which serves as the obstacle, was artificially created by focusing and increasing the laser intensity beyond the sample damage threshold. We find that at polariton densities of approximately $0.5 \cdot 10^6~\text{pol}/\upmu \text{m}^2$, the defect causes elastic scattering of polaritons, which can be observed as a bright Rayleigh scattering ring in momentum space (Fig.~2c) and a modulation in real space characteristic of supersonic flow. The scattered polaritons interfere with the incoming flow, generating the spatial pattern of fringes visible in Fig.~2a. In the wake of the artificial defect a shadow cone appears due to the reduced transmission across the defect. However, when the pump power is increased such that the polariton density reaches values close to $10^7~\text{pol}/\upmu \text{m}^2$, the shadow cone as well as the waves upstream vanish almost completely, as expected for a superfluid. This is confirmed by the image in momentum space (Fig.~2d), in which the scattering ring disappears almost fully. Note that some weak normal flow and scattering is perceivable in Figs.~2b,d. This is due to the tail ends of the polariton pulse before and after the superfluidity threshold, which are captured by the time-integrated images.

\begin{figure}[htbp]
  \centering \includegraphics[width=15cm]{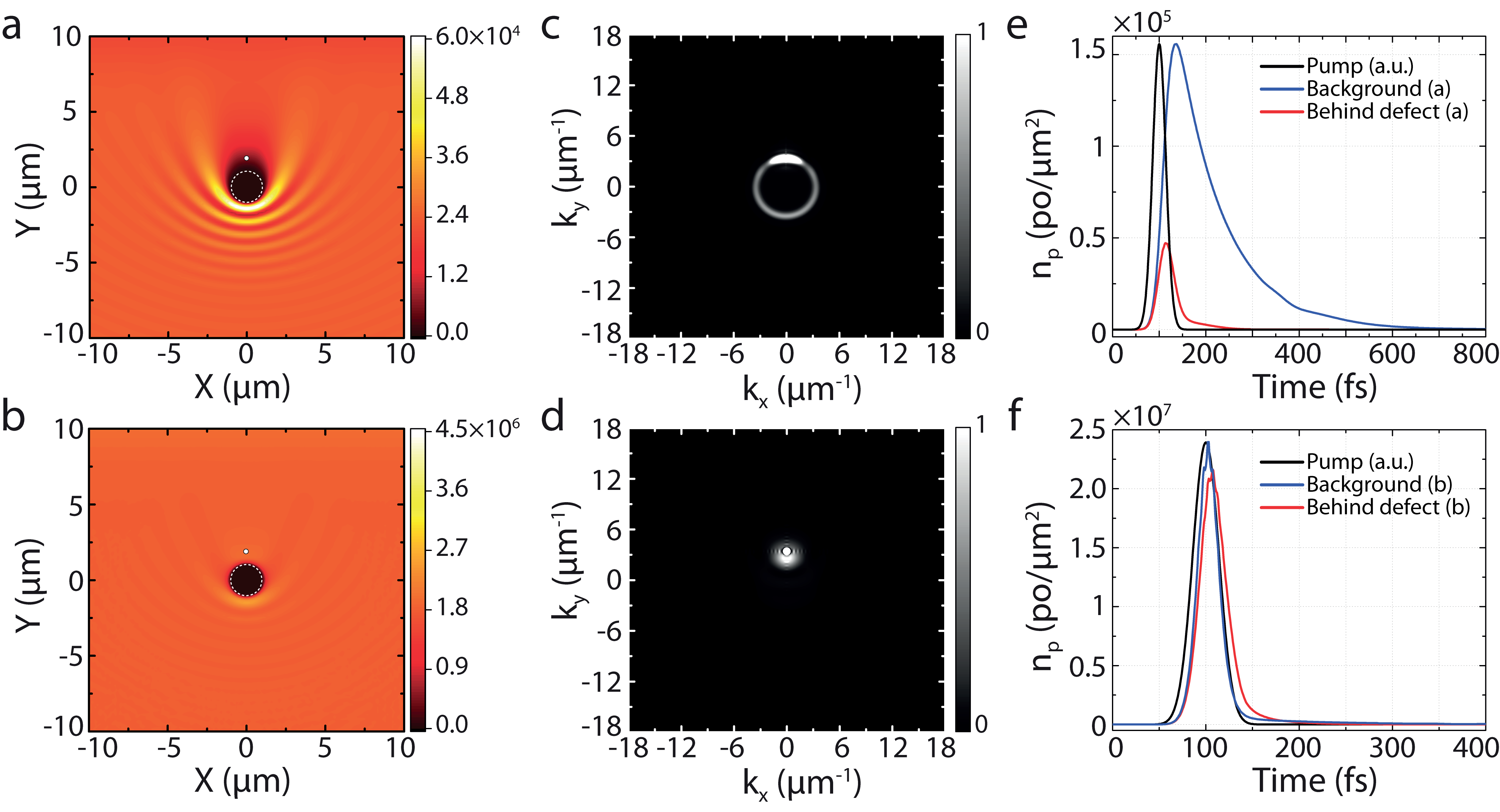} \linespread{1.1}
  \protect\protect\caption{
\textbf{Numerical calculations.} Real space polariton density profiles in the \textbf{a,} linear and \textbf{b,} superfluid regimes with peak background densities of $1.6 \cdot 10^5~\text{pol}/\upmu \text{m}^2$ and $2.4 \cdot 10^7~\text{pol}/\upmu \text{m}^2$, respectively. Dashed circle illustrates the position of a circular defect with radius $1~\upmu \text{m}$. The white dot is a probe spot placed $0.5~\upmu \text{m}$ behind the defect. Color scale is in $\text{pol}/\upmu \text{m}^2$. Saturated intensity momentum-space density profiles showing \textbf{c,} the presence of the Rayleigh scattering ring pattern in the linear regime and \textbf{d,} its suppression in the superfluid regime. Time-domain polariton density traces behind the defect (red) and in the background (blue) in the \textbf{e,} linear and \textbf{f,} superfluid regimes. The pump pulse is traced in black and normalized to the peak background density.
  }
\label{fig:FIG3} 
\end{figure}

By solving the time-dependent Gross-Pitaevskii equation, we can calculate the polariton spatial distribution using parameters corresponding to our experimental conditions (see Methods). Results for the linear (superfluid) regime are shown in the top (bottom) row of Fig.~3. To compare with the experimental time-integrated images, Figs.~3a and 3b show the calculated polariton density $|\psi(\vec{r},t)|^2$ integrated over a time interval of $800~\text{fs}$. The dashed circle indicates the location of a circular defect of radius $1~\upmu \text{m}$ and, consistently with the experimental images, the polariton flow is oriented in the positive $\hat{y}$  direction. Figure 3a shows the linear regime with a peak polariton density of $1.6 \cdot 10^5~\text{pol}/\upmu \text{m}^2$, where the polariton-polariton interaction energy $g|\psi(\vec{r},t)|^2$ remains small compared to the other terms in the GP equation. The calculation reproduces the characteristic features of the experiment (Fig.~2a), namely a parabolic modulation of the polariton density in front of the defect, resulting from interference between incident and scattered waves, and a shadow or decreased density behind the defect. Figure 3b shows the case where the incident fluence is increased such that the peak polariton density reaches $2.4 \cdot 10^7~\text{pol}/\upmu \text{m}^2$. Here, scattering by the defect is suppressed, which corresponds to the superfluid regime. As in the experiment (Fig.~2b), the interference fringes around the defect remain slightly visible due to the inclusion of time periods with polariton densities below the superfluid threshold.

The time-integrated momentum-space densities are shown in Figs.~3c and 3d. In Fig.~3c, which corresponds to the linear regime, the saturated intensity scale has been adjusted to highlight the Rayleigh scattering ring created by the scattering of polaritons to isoenergetic states possessing the same wavevector magnitude. In strong contrast, Fig.~3d shows the superfluid momentum space image where the Rayleigh scattering ring collapses to a small region around the incident wavevector.

The time-evolution of the polariton densities taken behind the defect (white dot in Figs.~3a and 3b) and in the background far away from the defect are shown in Figs.~3e and 3f for the linear and superfluid regimes, respectively. For reference, the normalized pump pulse is shown in black. In the linear regime, we can observe the reduced polariton density behind the defect and a slow decay of the polariton densities on a timescale given by the polariton lifetime. As the pump density is increased, we first cross into a regime where the driving rate exceeds dissipation and the field coherently follows that of the pump. This is observed as a faster decay of the polariton density, which tracks the pump, but is not sufficient to observe superfluidity in the absence of interactions (Fig.~S2). Beyond this threshold, we cross into the superfluid regime due to the renormalized dispersion. Here, the density behind the defect follows the background density closely, with the exception of some damped oscillations after the end of the excitation pulse. When crossing back into dissipative flow at longer times, vortex pairs are created behind the defect and carried away by the viscous flow (Figs.~S3-S4). In the pulsed experiment, the normal regime is always recovered within the tail end of the pump and the steady-state dynamics, which would occur at longer times, are not directly accessible.  We have, however, performed time-domain simulations to highlight the anticipated behaviour (Fig.~S5).

\begin{figure}[htbp]
  \centering \includegraphics[width=10cm]{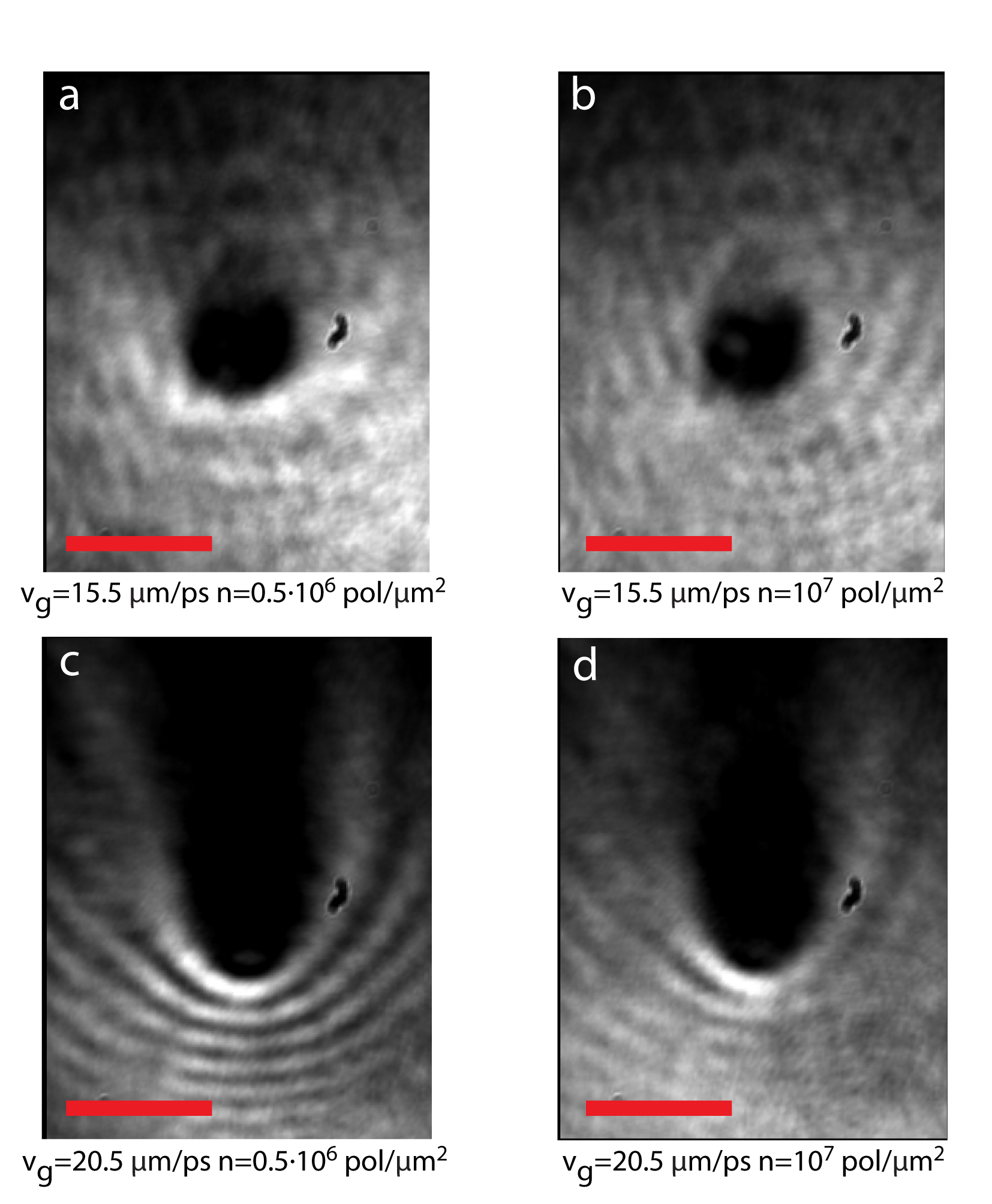} \linespread{1.1}
  \protect\protect\caption{
\textbf{Changing the polariton group velocity.}  Real space intensity distribution of the polariton flux (flowing from the bottom to the top of the images) across the same artificial defect reported in Fig.~2. The polariton group speed is reduced down to $15.5~\upmu \text{m/ps}$ in \textbf{a,b} and increased up to $20.5~\upmu \text{m/ps}$ in \textbf{c,d} compared to the value in Fig.~2. However, the polariton densities are kept the same: $0.5 \cdot 10^6~\text{pol}/\upmu \text{m}^2$ (a,c) and $10^7~\text{pol}/\upmu \text{m}^2$ (b,d). In a, the image shows a condition closer to the superfluid threshold than that in Fig.~2a, due to the reduced speed of the fluid. It shows a complete transition to the superfluid regime for higher densities (b). At higher group velocities, for low polariton densities, the system is fully in the supersonic regime, clearly evidenced by the fringes and the shadow cone in c. However for such high velocity even at the highest densities the fluid  still remains supersonic (d). These highlight the important role of the polariton velocity on achieving superfluid behaviour. The scale bar is $5~\upmu \text{m}$.
 }
\label{fig:FIG4} 
\end{figure}

To test the effect of the polariton group velocity on the superfluid behaviour, we have repeated the experiments shown in Fig.~2, but for different resonant excitation conditions along the polariton dispersion. When the group velocity is lowered to $15.5~\upmu \text{m/ps}$, the interference fringes in the linear regime ($0.5 \cdot 10^6~\text{pol}/\upmu \text{m}^2$) are less evident than in the previous case (Fig.~4a). This effect is caused by the small propagation length of the fluid at this wavevector, but also to a lowering of the superfluid threshold. The density distribution shows a higher fluid density immediately ahead of the defect, similar to the case of atomic condensates when perturbed by a laser field scanned above the critical velocity~\cite{onofrio_observation_2000}. Indeed, when the density is increased ($10^7~\text{pol}/\upmu \text{m}^2$), the superfluid regime is recovered: the intensity peak before the defect disappears and the spatial distribution of the population becomes homogeneous around the defect (Fig.~4b). On the contrary, when the polariton speed is increased to $20.5~\upmu \text{m/ps}$, superfluid behaviour is no longer achievable in the range of polariton densities attainable below the damage threshold of our sample. Here, the waves and shadow cone always remain visible in the images (Figs.~4c and 4d).

\begin{figure}[htbp]
  \centering \includegraphics[width=16cm]{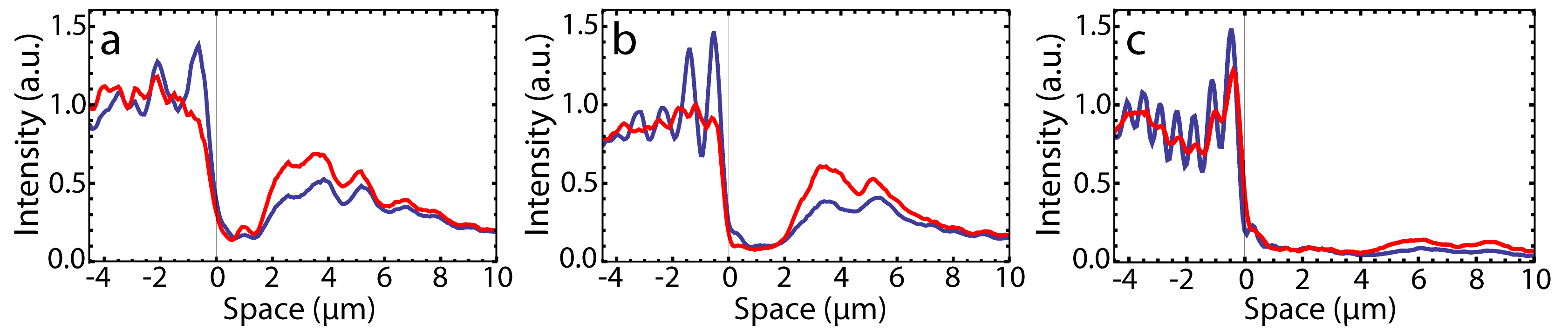} \linespread{1.1}
  \protect\protect\caption{
\textbf{Intensity profiles.} \textbf{a,} The blue line shows the intensity profile of the time-integrated propagating polariton wavepacket hitting a defect at a polariton density of $0.5 \cdot 10^6~\text{pol}/\upmu \text{m}^2$ and group velocity of $15.5~\upmu \text{m/ps}$. Increasing the polariton density to $10^7~\text{pol}/\upmu \text{m}^2$ (red line), the spatial modulation of the polariton population close to defect position is suppressed and the intensity profile after the defect recovers the shape of a propagating Gaussian wavepacket with a finite lifetime of $100~\text{fs}$. \textbf{b,} As in panel a. but with a polariton group velocity of $19~\upmu \text{m/ps}$. The higher velocity affects the modulation frequency of the spatial waves before the defect (blue line). At the highest polariton densities (red line) superfluidity is still recovered. \textbf{c,} When the polariton group velocity is increased up to $20.5~\upmu \text{m/ps}$, both the modulation and the shadow cone are present even for the highest polariton density of $10^7~\text{pol}/\upmu \text{m}^2$ (red line).
  }
\label{fig:FIG5} 
\end{figure}

Figure 5 shows vertical line profiles taken along the central axis of the defect for the three different group velocities reported above. When the group velocity increases (i.e. from panel a to panel c) at a polariton density of $0.5\cdot 10^6~\text{pol}/\upmu \text{m}^2$ (blue lines), the interference fringes generated by elastic scattering increase in spatial frequency. The spatial modulation of the polariton population, caused by the presence of the defect along the wavepacket path, vanishes only when the polariton densities are increased to $10^7~\text{pol}/\upmu \text{m}^2$ (red lines) and for group velocities below $19~\upmu \text{m/ps}$ (Fig.~5a,b). In fact, both the density peak immediately before and the shadow cone in the wake of the defect persist at a velocity of $20.5~\upmu \text{m/ps}$ for a polariton density of $10^7~\text{pol}/\upmu\text{m}^2$ (Fig.~5c). Note that the increase of the group velocities in the supersonic regime results in a higher modulation of the interference fringes ahead of the obstacle (see blue lines in Fig.~5).

Similar results have been obtained by measuring scattering from natural defects formed during the microcavity fabrication process as shown in Fig.~S6. However, in this case, the shallowness of the defect and its wider extent make the visibility of the fringes as well as the presence of the shadow cone of much lesser sharpness. Yet even in this case a smooth transition to a superfluid regime, at high polariton densities, can be perceived.\\

\noindent
\textbf{Conclusions\\}
We have experimentally demonstrated the superfluid flow of polaritons around a defect in an organic microcavity under ambient conditions. Strikingly, the perturbations induced by artificial and natural potential barriers on the polariton flow vanish for high enough particle densities and low group velocities. In agreement with the experimental results, GP calculations show that despite the small polariton-polariton interaction constant of organic polaritons, the superfluid regime at group velocity of $19~\upmu \text{m/ps}$ is achieved when the polariton density is increased to approximately $10^7~\text{pol}/\upmu\text{m}^2$.  To our knowledge, this is the first time that such superfluid behaviour has been observed at room temperature and this has important implications for the realization of photonic devices protected from scattering by means of polariton-polariton interactions. It is worth noting that in polariton condensates under resonant excitation, a definition of the superfluid fraction, which is the part of the fluid that responds only to irrotational (non-transverse) forces, is still an open and important question\cite{keeling_superfluid_2011}. In this context, the short pulse excitation used in the present experiment is particularly useful, because it could allow for spatially separated excitation in samples with longer polariton lifetimes. On the other hand, the intermediate pulse length regime is also of interest as it would allow for the observation of quasi steady-state behavior free of transients and relaxation oscillations at early times. Under this configuration, single-shot measurements could allow the observation of free vortex formation for suitable conditions of the barrier potential and fluid density, opening the way to the study of hydrodynamic effects in quantum fluids on room-temperature table-top experiments.\\

\noindent
\textbf{Methods}

\noindent
\textbf{Experiment\\}
The microcavity under investigation is similar to that in Ref.~\citenum{daskalakis_spatial_2015} and consists of a 130 nm thick film of layer of TDAF sandwiched between two dielectric Bragg reflectors.  The organic layer was thermally evaporated in vacuum and the mirrors, which are composed of 9 pairs of alternating $\text{Ta}_2\text{O}_5/\text{SiO}_2$, sputtered in the same chamber.
Tunable short pulse excitation was obtained using an optical parametric amplifier pumped by a Ti:Sapphire regenerative amplifier at $10~\text{kHz}$. Polaritons with a specific group velocity were created by focusing the laser on the back focal plane of the first microscope objective, which was a $60\times$ long working distance objective on the substrate side. The overall magnification for real space in detection was $300\times$ and the excitation spot FWHM was approximately $13~\upmu \text{m}$. Furthermore, we use a white light source incident from the detection side (in the reflectance configuration) to measure the polariton dispersion, facilitating resonant tuning of the excitation laser on the polariton branch. The FWHM of the excited momenta is $0.55~\upmu \text{m}^{-1}$ and the incident laser spectral width corresponds to a FWHM of $7.8~\text{nm}$ at $\lambda = 424~\text{nm}$.
In all of the experiments, care is taken to ensure that after reaching the highest polariton density the initial condition is fully recovered after reducing the pump power to its initial value. This limits the incident fluence to approximately $4~\text{mJ}/\text{cm}^2$ and short measurement intervals.\\

\noindent
\textbf{Gross-Pitaevskii simulations\\}
Because of the small interaction energy, as compared to the splitting between lower (LP) and upper polariton branches, it is sufficient to consider only the lower polariton field $\psi(\vec{r},t)$. Its time-evolution is governed by a generalized Gross-Pitaevskii equation:
\begin{equation}
i\hbar \frac{\partial\psi(\vec{r},t)}{\partial t}= \left[ \hbar \omega_0 - \frac{\hbar^2 \nabla^2}{2m_{LP}} + V(\vec{r}) - \frac{i\hbar \gamma_{LP}}{2} + g|\psi(\vec{r},t)|^2\right]\psi(\vec{r},t) + \hbar P(\vec{r},t) 
\end{equation}
The first two terms on the right-hand side correspond to a parabolic approximation for the lower polariton branch dispersion, which is valid under our experimental conditions. Here $\hbar \omega_0$ is the LP energy at $k = 0$, $m_{LP}$ is the effective mass, $V(\vec{r})$ is the scattering defect potential, $\gamma_{LP}$ is the lower polariton dissipation rate, $g$ is the polariton-polariton interaction constant and $P(\vec{r},t)$ is the driving term. 
The pump field is taken to be a plane wave modulated by a temporal Gaussian envelope
\begin{equation}
P(\vec{r},t) = F_p e^{i( \vec{k} \cdot \vec{r} - \omega_p t)} e^{-\frac{(t-t_0)^2}{2\sigma_t^2}} 
\end{equation}
where $\sigma_t$ can be related to the intensity FWHM using $\sigma_t=FWHM/(2\sqrt{ln2})$. The possible influence of chirp has been considered in the Supplementary Information and was not found to qualitatively affect the behaviour. The amplitude  of the driving term can be related to the incident pump intensity using input-output theory~\cite{wouters_parametric_2007} as
\begin{equation}
F_p = C_{k_p} \sqrt{\frac{\gamma_{LP} I_0}{2\hbar \omega_p}}
\end{equation}
where $C_{k_p} = 0.88$ is the Hopfield coefficient for the photon fraction of the lower polariton branch at the pump wavevector $k_p$. For simplicity, the defect is taken to be an infinite barrier with vanishing boundary conditions. The remaining simulation parameters are $\hbar \omega_0 = 2.896~\text{eV}$, $m_{LP} = 1.976 \cdot 10^{-35}~\text{kg}$, $\gamma_{LP} = 10^{13}~\text{s}^{-1}$, $g = 5 \cdot 10^{-3}~\upmu \text{eV} \cdot \upmu \text{m}^2$, $\vec{k}_p = 3.59~\hat{y}~\upmu \text{m}^{-1}$ and $\hbar \omega_p = 2.9242~\text{eV}$. 

To make a connection to the pump fluence, we can use $E_{in} = I_0 \sigma_t \sqrt{\pi}$, which gives $26.6~\text{mJ}/\text{cm}^2$ for the incident fluence in the superfluid regime. This is comparable with the experimental value of $4~\text{mJ}/\text{cm}^2$,  given the approximations made in connection with input-output theory and the approximate value of the polariton interaction constant. Indeed, the latter is the only experimental parameter not known with certainty. The value chosen here is close to our previous estimate of $g = 10^{-3}~\upmu \text{eV} \cdot \upmu \text{m}^2$ in this structure~\cite{daskalakis_spatial_2015} and also to that obtained using the resonant blueshift at high powers (see Supplementary Information). At higher powers, however, the measured blueshift of the polariton may be strongly affected by the thermal load of the pump, causing a competing redshift of the polariton energy due to an expansion of the cavity length. Another red-shift contribution may also arise from polaron-polariton coupling~\cite{wu_when_2016} at these fluences.



\newpage

\noindent
\textbf{Acknowledgements} This work was funded by the ERC project POLAFLOW (grant no.~308136). F.B. and S.K.C. acknowledge funding from the NSERC Discovery Grant and the Canada Research Chair program. S.A.M acknowledges the Leverhulme Trust and EPSRC Active Plasmonics Programme and K.S.D. acknowledges funding from the Academy of Finland through its Centers of Excellence Programme (2012-2017) under project No. 284621 and the European Research Council (ERC-2013-AdG-340748-CODE).\\

\noindent
\textbf{Author contributions}
G.L. conceived and performed the optical measurements with assistance from A.F. K.S.D. fabricated the sample and F.B. performed the simulations. G.L., D.B., F.B., K.S.D., S.K.C., D.S. co-wrote the manuscript. All authors contributed to the data analysis. S.K.C. supervised the fabrication and simulation work and D.S. supervised the measurements and coordinated the project. \\

\noindent
\textbf{Competing interests} The authors declare that they have no competing financial interests.\\

\noindent
\textbf{Correspondence} Correspondence and requests for materials should be addressed to D.B.~(email: dario.ballarini@gmail.com).

\newpage
\onecolumngrid

\setcounter{equation}{0}
\setcounter{figure}{0}
\setcounter{table}{0}
\setcounter{page}{1}
\makeatletter
\renewcommand{\theequation}{S\arabic{equation}}
\renewcommand{\thefigure}{S\arabic{figure}}
\pagenumbering{roman}

\noindent
\textbf{\Large Supporting Information}\\

\noindent
\textbf{a) Steady-state Bogoliubov excitation spectra}

The excitation spectra of the stationary solutions can be calculated by linearizing the GP equation around a stationary homogeneous state $\psi_{SS}$ following Ref.~\citenum{carusotto_probing_2004}. We consider the case of resonant excitation at the renormalized dispersion and that the LP wavefunction has a plane-wave form similar to the pump field given by 
\begin{equation}
\psi(\vec{r},t) = \psi_{SS} e^{i( \vec{k}_p \cdot \vec{r} - \omega_p t)}  
\end{equation}
Then, the spectrum of the excitations is given by the solutions of the eigenvalue problem 
\begin{equation}
\mathcal{L}_k \mathcal{U}_k = \hbar \omega_k \mathcal{U}_k
\end{equation}
with the small fluctuations vector given by
\begin{equation}
\mathcal{U}_k = [ \delta \psi(\vec{r},t) , \delta \psi^*(\vec{r},t) ]^T
\end{equation}
and the operator $\mathcal{L}_k$ being defined as
\[
\mathcal{L}_k= 
\begin{pmatrix}
\omega_{LP}(k) + 2gn_p - i\gamma_{LP}/2 & gn_p \\
-gn_p & 2\omega_p - \omega_{LP}(2k_p-k) - 2gn_p - i\gamma_{LP}/2
\end{pmatrix}
\]
Figure S1 shows the Bogoliubov excitation energies obtained for polariton density values below, at and above the superfluidity threshold. The energies are traced with respect to the center pump energy. Note that for each value of $k$, the dispersion contains two branches $\omega_{LP}^{\pm}(k)$, traced in red ($+$) and blue ($-$), that are related by symmetry as 
$\omega_{LP}^{-}(k)=2\omega_p - \omega_{LP}^{+}(2k_p - k)$. 
At low densities, as in Fig.~S1a, the dispersion remains parabolic and the presence of isoenergetic states available for scattering generates the resonant Rayleigh scattering ring pattern observed in momentum space in Figs.~2c and 3c. For increasing densities, the polariton-polariton interactions tilt the dispersion and introduce a discontinuity in the slope around $k_p$. For the threshold density shown in Fig.~S1b, the two dispersions are horizontal at low momentum values in the vicinity of $k_p$ and point to the collapse of the scattering ring to a single point, as shown in Figs.~2d and 3d. Above the superfluidity threshold, as in Fig.~S1c, the two dispersions only touch at $k_p$ and the linear slope defines the sound velocity. The threshold density for the superfluid transition is in agreement with the experimental densities and the time-dependent calculations.\\

\begin{figure}[htbp]
  \centering \includegraphics[width=16cm]{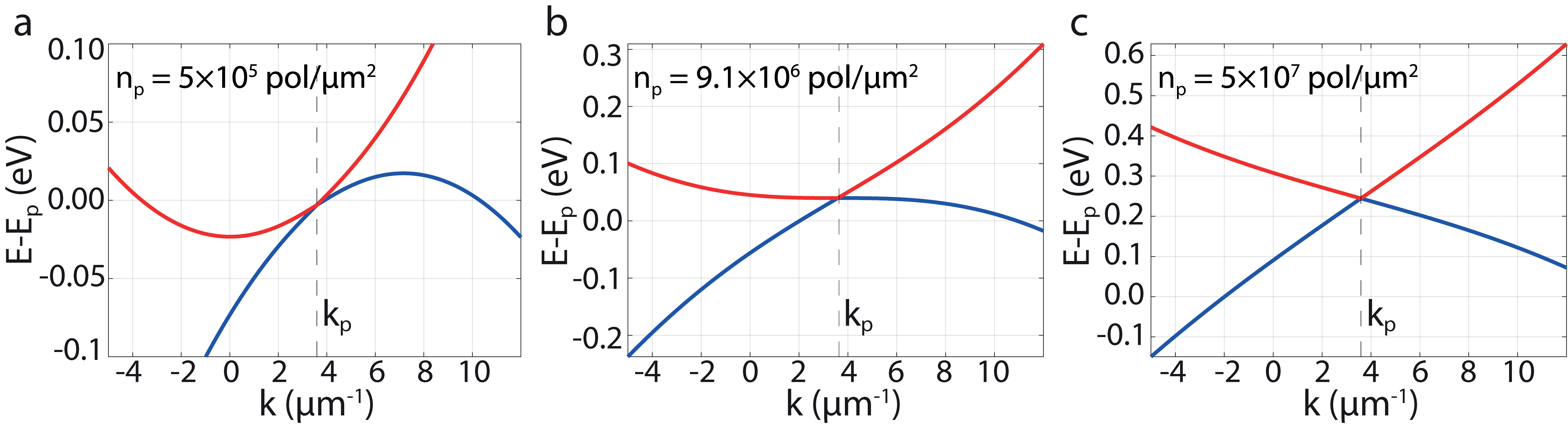} \linespread{1.1}
  \protect\protect\caption{
\textbf{Bogoliubov excitation spectra for the LP branch.} Positive (red) and negative (blue) LP Bogoliubov modes, calculated in steady-state and at resonance with the renormalized LP dispersion, traced with respect to the center pump energy. The pump wavevector is indicated by a dashed vertical line. \textbf{a,} For low densities, the dispersion remains parabolic. \textbf{b,} Increasing the density tilts the dispersion and introduces a discontinuity in the slope and, at threshold, the curves are horizontal for low momentum values around the pump wavevector. \textbf{c,} Above the superfluidity threshold, the dispersions only touch at the pump wavevector. Resonant Rayleigh scattering is no longer possible and the polariton fluid displays superfluid behaviour.\\
 }
\label{fig:FIGS1} 
\end{figure}

\noindent
\textbf{b) Superfluid pump fluence with interaction constant $g$ set to zero}

In order to verify that the suppression of scattering in the time integrated images was not due to the shortening in time of the polariton density under high excitation, the interaction constant $g$ was set to zero while keeping the remaining simulation parameters for the superfluid regime unchanged. As shown in Fig.~S2, the results obtained in this case are identical to the ones shown in Fig.~3a for the linear regime, but with an increased density.

\begin{figure}[htbp]
  \centering \includegraphics[width=16cm]{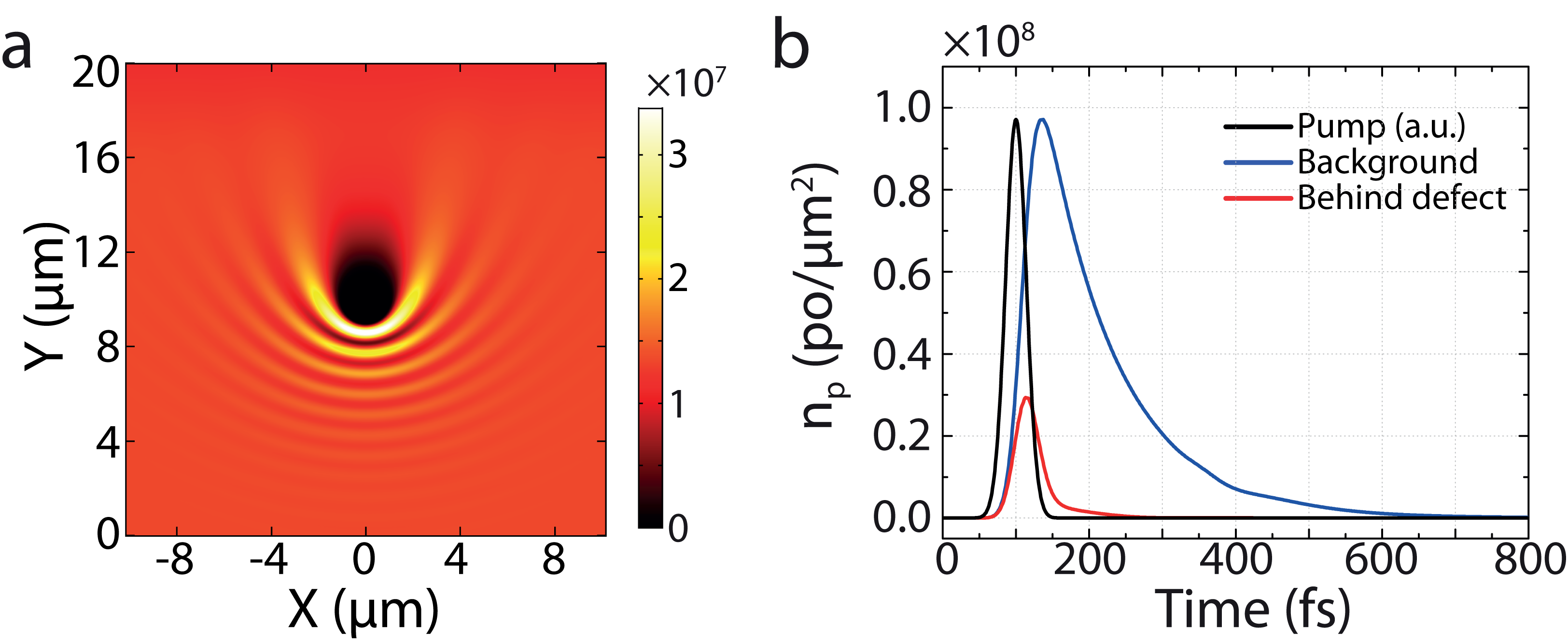} \linespread{1.1}
  \protect\protect\caption{
\textbf{Superfluid case with interaction constant $g$ set to zero.} \textbf{a,} Real space polariton density profile with the same parameters used in the superfluid regime (Fig.~3d) but with the interaction constant $g$ set to zero. Peak background density is $9.7 \cdot 10^7~\text{pol}/\upmu\text{m}^2$. \textbf{b,} Time-domain polariton density traces behind the defect (red) and in the background (blue). The pump pulse is traced in black and normalized to the peak background density.
 }
\label{fig:FIGS2} 
\end{figure}

\noindent
\textbf{c) Time evolution}

Figures S3 and S4 show instantaneous snapshots of the calculated polariton densities in the linear and superfluid regimes, respectively.

\begin{figure}[htbp]
  \centering \includegraphics[width=16cm]{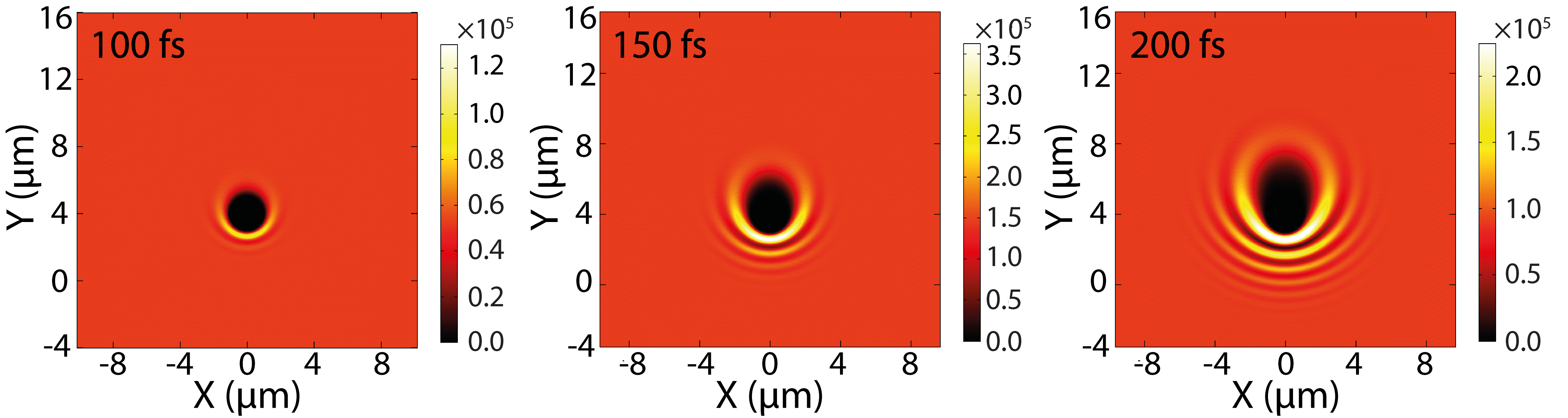} \linespread{1.1}
  \protect\protect\caption{
\textbf{Time-evolution of polariton density in the linear regime.} Individual time snapshots of the polariton density during polariton flow. The color scales are in $\text{pol}/\upmu\text{m}^2$ and their maximum values are adjusted to maintain the background color throughout the images.
 }
\label{fig:FIGS3} 
\end{figure}

\begin{figure}[htbp]
  \centering \includegraphics[width=16cm]{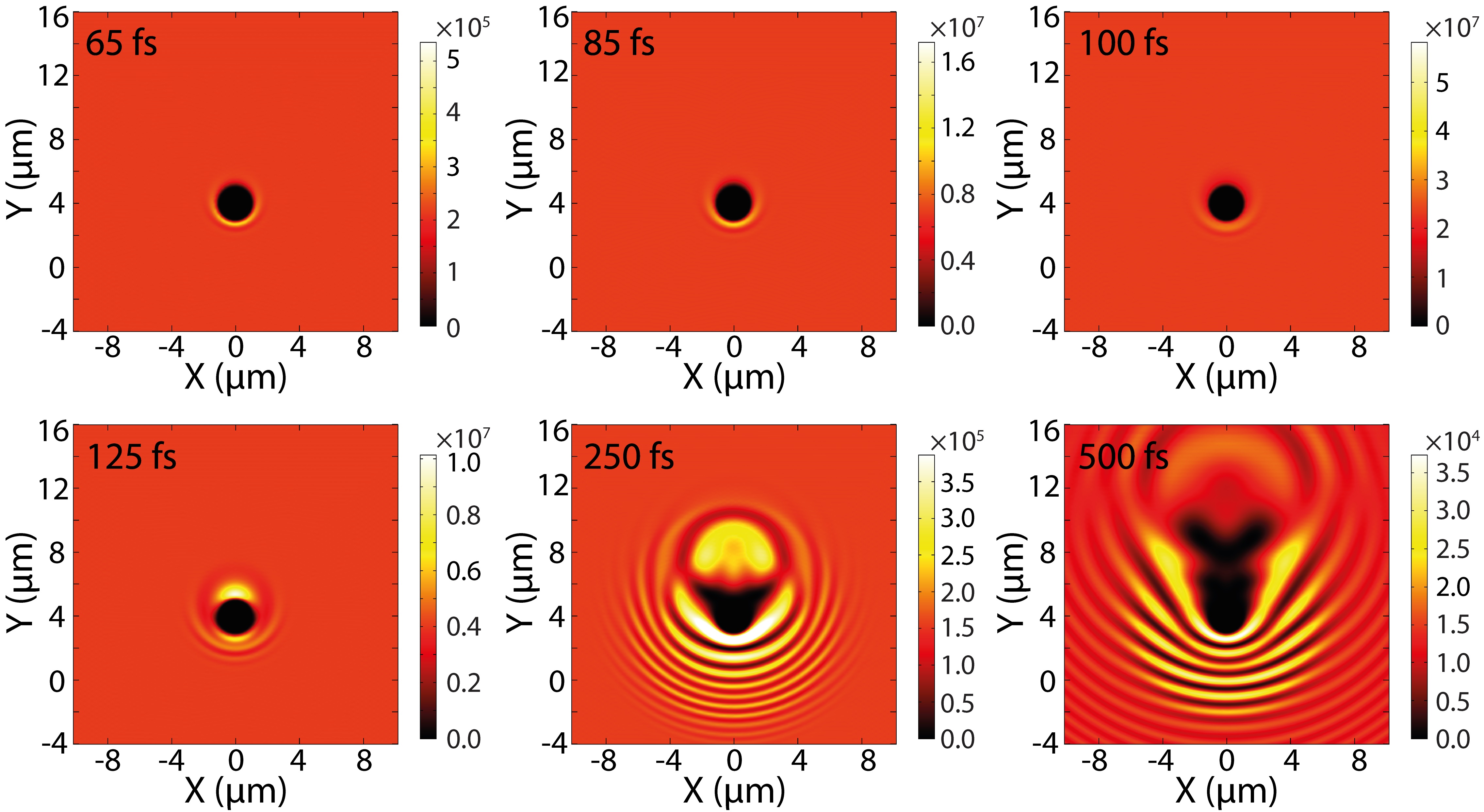} \linespread{1.1}
  \protect\protect\caption{
\textbf{Time-evolution of polariton density in the superfluid regime.} Individual time snapshots of the polariton density during polariton flow. Note the ejection of vortex pairs around $500~\text{fs}$ when the density has been reduced. The color scales are in $\text{pol}/\upmu\text{m}^2$ and their maximum values are adjusted to maintain the background color throughout the images.
 }
\label{fig:FIGS4} 
\end{figure}

\noindent
\textbf{d) Steady-state}

To observe the steady-state behaviour of the superfluid case, the pump pulse was set to be a long rectangular pulse with duration of $2~\text{ps}$, having the same peak amplitude as in the short pulse case shown in Fig.~3d. As shown in Fig.~S5, superfluidity persists through the damped oscillations and the linear regime is recovered shortly after the end of the pump pulse.

\begin{figure}[htbp]
  \centering \includegraphics[width=16cm]{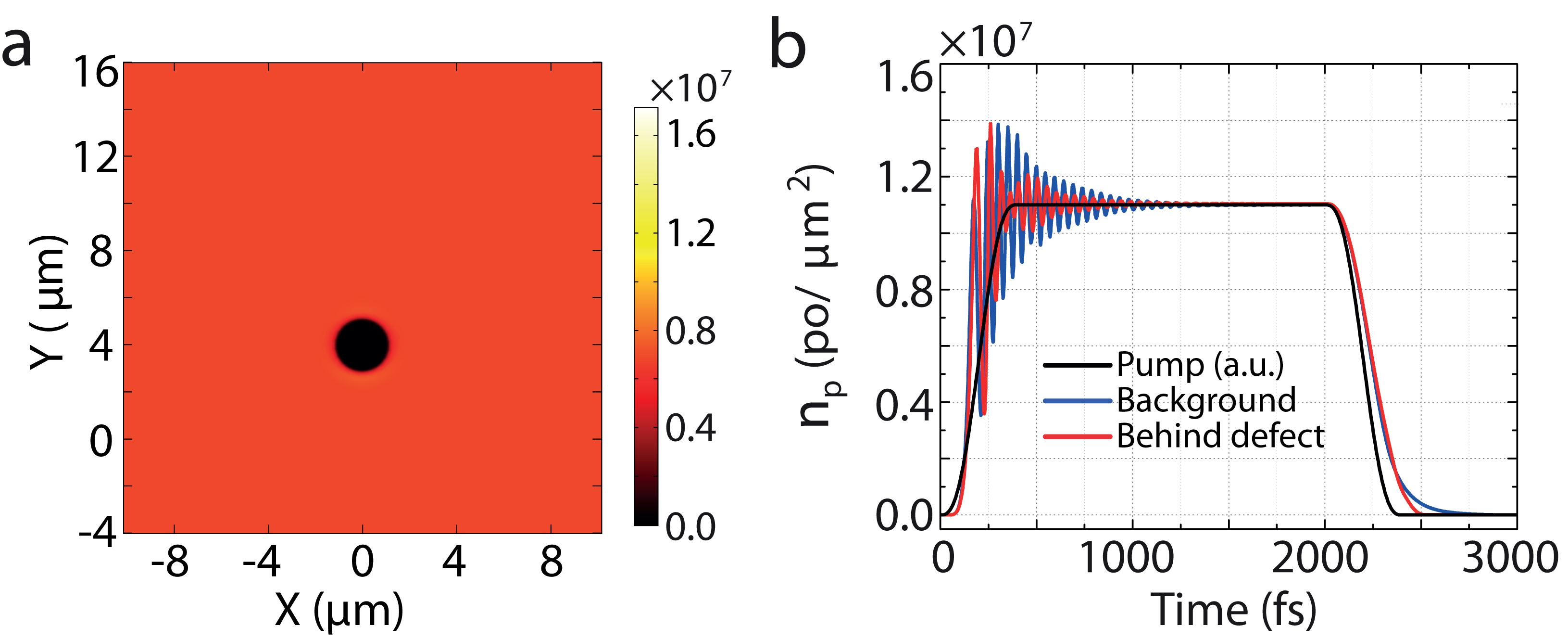} \linespread{1.1}
  \protect\protect\caption{
\textbf{Steady-state flow.} \textbf{a,} Real space polariton density profile in the superfluid regime when the pump is a rectangular pulse of duration $2~\text{ps}$. Steady-state background density is  $1.1 \cdot 10^7~\text{pol}/\upmu\text{m}^2$. \textbf{b,} Time-domain polariton density traces behind the defect (red) and in the background (blue). The pump pulse is traced in black and normalized to the steady-state density.
 }
\label{fig:FIGS5} 
\end{figure}

\noindent
\textbf{e) Superfluid flow around a natural defect}

\begin{figure}[b]
  \centering \includegraphics[width=15cm]{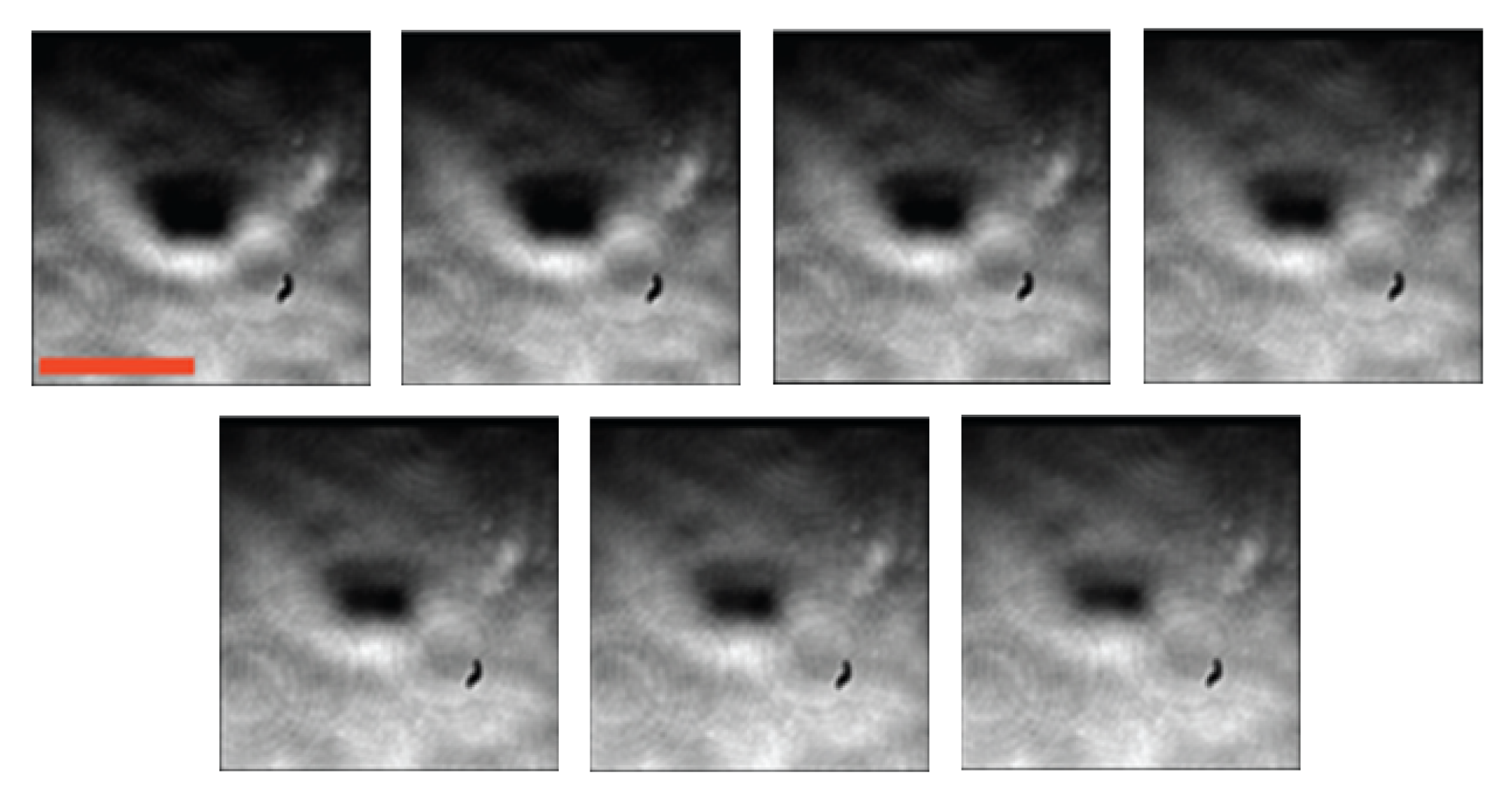} \linespread{1.1}
  \protect\protect\caption{
\textbf{Natural defect.} Real space imaging of the polariton flux with increasing polariton densities when hitting a natural defect on the thin film deposition. The scale bar is $5~\upmu\text{m}$. The polariton densities are $0.027,~0.074,~0.2,~0.55,~1.48,~4,~11$ ($10^6~\text{pol}/\upmu\text{m}^2$) from left-to right, top to bottom.
 }
\label{fig:FIGS6} 
\end{figure}

A polariton wavepacket travelling through a defect spontaneously formed during the sample fabrication process is injected into the microcavity at a group velocity of $19~\upmu \text{m/ps}$. Figure S6 shows the density map in space at different pumping powers. A smooth crossover to the superfluid behaviour is visible when increasing the polariton density. However, because of the lack of control over the shape and depth of the natural defects, the superfluidity behaviour is less evident than in artificial defects. Indeed, in this latter case, the geometrical parameters are easy tunable allowing the perfect visibility of the interference fringes and the shadow cone in the wake of the defect, as shown in the figures of the main text.\\

\noindent
\textbf{f) Interaction constant}

\begin{figure}[b]
  \centering \includegraphics[width=11cm]{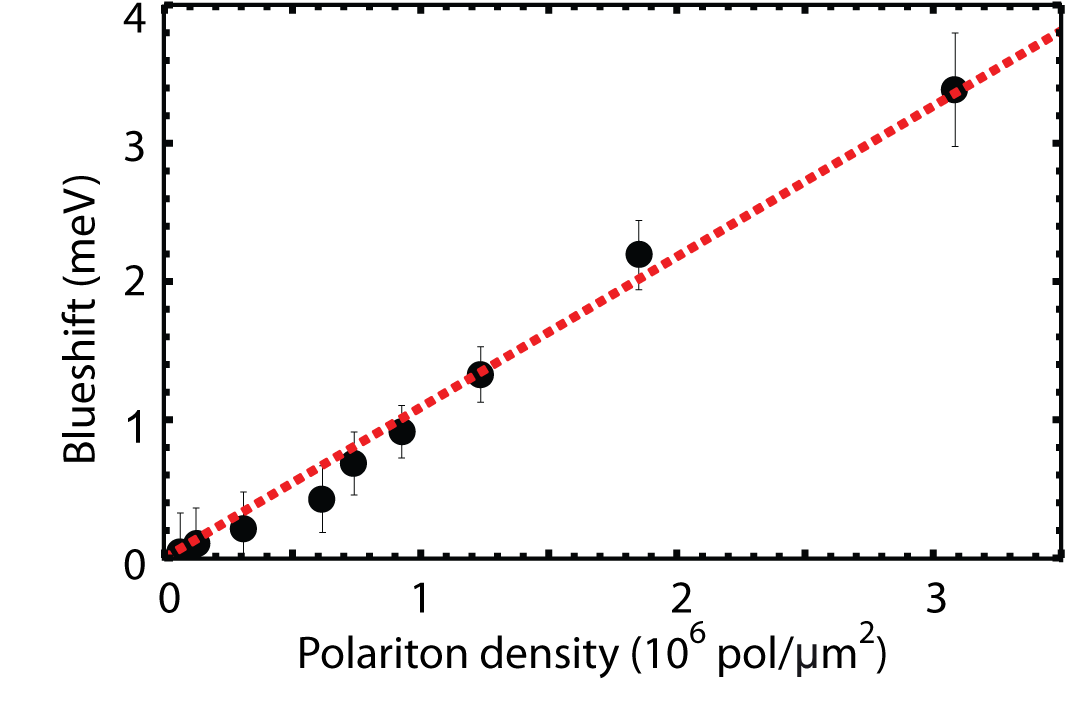} \linespread{1.1}
  \protect\protect\caption{
\textbf{Interaction constant.} Experimental blueshift of the polariton resonance (black dots) through resonant pumping at $k = 0$. The red dashed curve is the linear fitting of the data, returning a polariton-polariton interaction constant $g = 10^{-3}~\upmu\text{eV} \cdot \upmu\text{m}^2$.
 }
\label{fig:FIGS7} 
\end{figure}

We attempted to estimate the interaction constant $g$ using resonant excitation of the LP branch. To avoid problems induced by polariton propagation outside the excitation area, the measurements  were performed by injecting polaritons with zero momentum. The excitation laser is the same used in the superfluidity experiments, set at $2.91~\text{eV}$ and with a FWHM of $7.8~\text{meV}$. The spot is focused on the sample with a Gaussian profile and a spatial FWHM of $5.78~\upmu \text{m}$. Figure S7 shows the polariton mode blueshift ($\Delta\text{E}$) at the center of the spot for different polariton densities. The interaction constant $g$ is evaluated fitting the spatial envelope of the polariton blueshift across the spot profile according to the relation $\Delta\text{E}(\vec{r}) = g|\psi(\vec{r})|^2$, where $|\psi(\vec{r})|^2$ is the polariton density at the spatial position $r$. The polariton density is estimated from the energy fluence transmitted through the cavity and assuming that half of the polaritons emit through the excitation side since the transmittance and reflectance values are both close to 50\% at the polariton resonance energy. We can infer from the experimental data a value of $g = 10^{-3}~\upmu\text{eV} \cdot \upmu\text{m}^2$.

\noindent
\textbf{g) Polariton propagation}

The polariton lifetime resulting from the mode energy linewidth ($7.3~\text{meV}$) is $90~\text{fs}$. Since this method does not consider linewidth broadening processes, it gives only a lower limit for the lifetime.
We can evaluate the true polariton lifetime by independent evaluations, i.e., by the propagation of a polariton wavepacket in the same configuration reported in the main text. Polaritons are injected with $19~\upmu \text{m/ps}$ group and the excitation spot is focused into a small Gaussian spot ($2~\upmu \text{m}$ FWHM) in order to clearly define its spatial exponential decay. The experimental data are fitted with the Gaussian profile of the spot convoluted with an exponential decay. The fitted propagation length is $1.9~\upmu \text{m}$, which corresponds to a polariton lifetime of $100~\text{fs}$, as shown in Fig.~S8.

\begin{figure}[htbp]
  \centering \includegraphics[width=12cm]{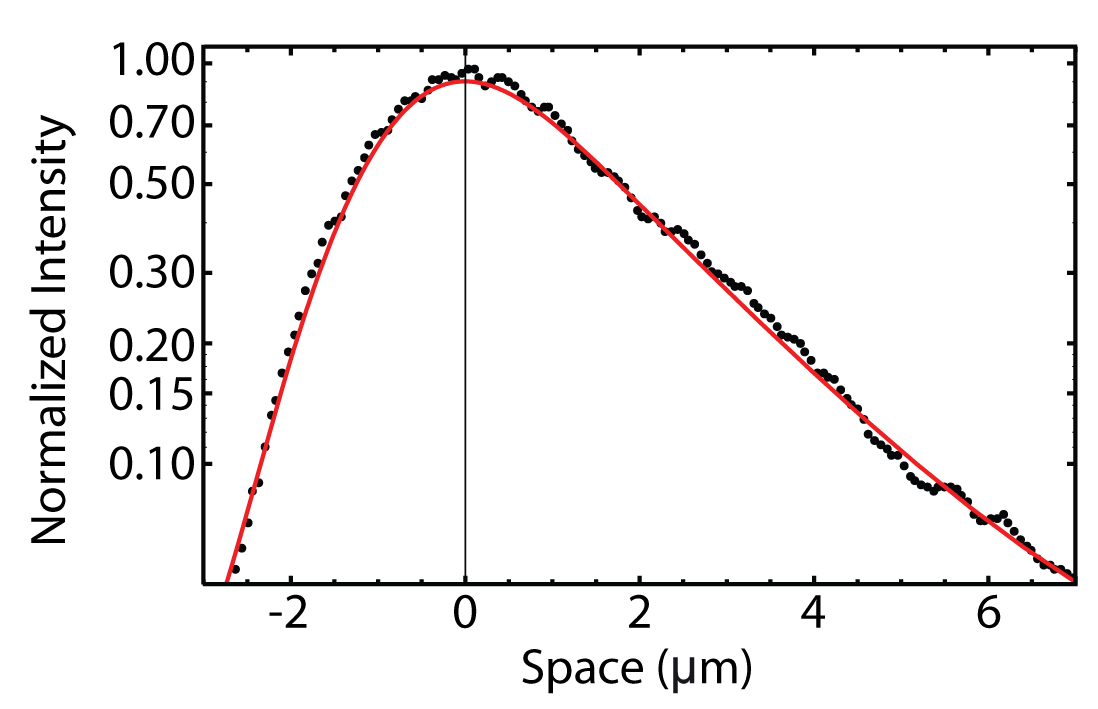} \linespread{1.1}
  \protect\protect\caption{
\textbf{Polariton propagation.} Intensity profile of a polariton wavepacket travelling at group velocity of $19~\upmu \text{m/ps}$. The best fitting of the experimental data (black dots) is obtained considering a Gaussian wavepacket with propagation length of $1.9~\upmu \text{m}$ (solid red line).\\ 
 }
\label{fig:FIGS8} 
\end{figure}

\noindent
\textbf{h) Temporal broadening of the pump pulse due to chirp}

The positive chirp acquired during propagation through optical elements leads to broadening of the pump pulse and lowering of its peak amplitude. To investigate the effects of chirp in the simulation results the equations of the driving term are modified as
\begin{equation}
P(\vec{r},t) = F_p e^{i( \vec{k} \cdot \vec{r} - \omega_p t)} e^{-\frac{(t-t_0)^2}{2 (\sigma_t^2 +  i \text{GVD} d)}} 
\end{equation}
and
\begin{equation}
F_p = C_{k_p} \sqrt{\frac{\gamma_{LP}I_0}{2\hbar\omega_p} \cdot \frac{\sigma_t^2}{\sigma_t^2 + i \text{GVD} d}}      
\end{equation}
where $\text{GVD} = 90.4~\text{fs}^2/\text{mm}$ is the group velocity dispersion coefficient of fused silica at the pump wavelength of $424~\text{nm}$ and $d = 3~\text{cm}$ is the total thickness of optical elements in the path of the pulse. The results for the nonlinear case are shown in Fig.~S9 for the same simulation parameters as in the bottom row of Fig.~3. The same superfluid behaviour is seen in the real and momentum space images, but the time domain traces show a a larger number of relaxation oscillations due to the longer pulse duration. The modulation from these initial oscillations is also larger than in the unchirped case.

\begin{figure}[htbp]
  \centering \includegraphics[width=16cm]{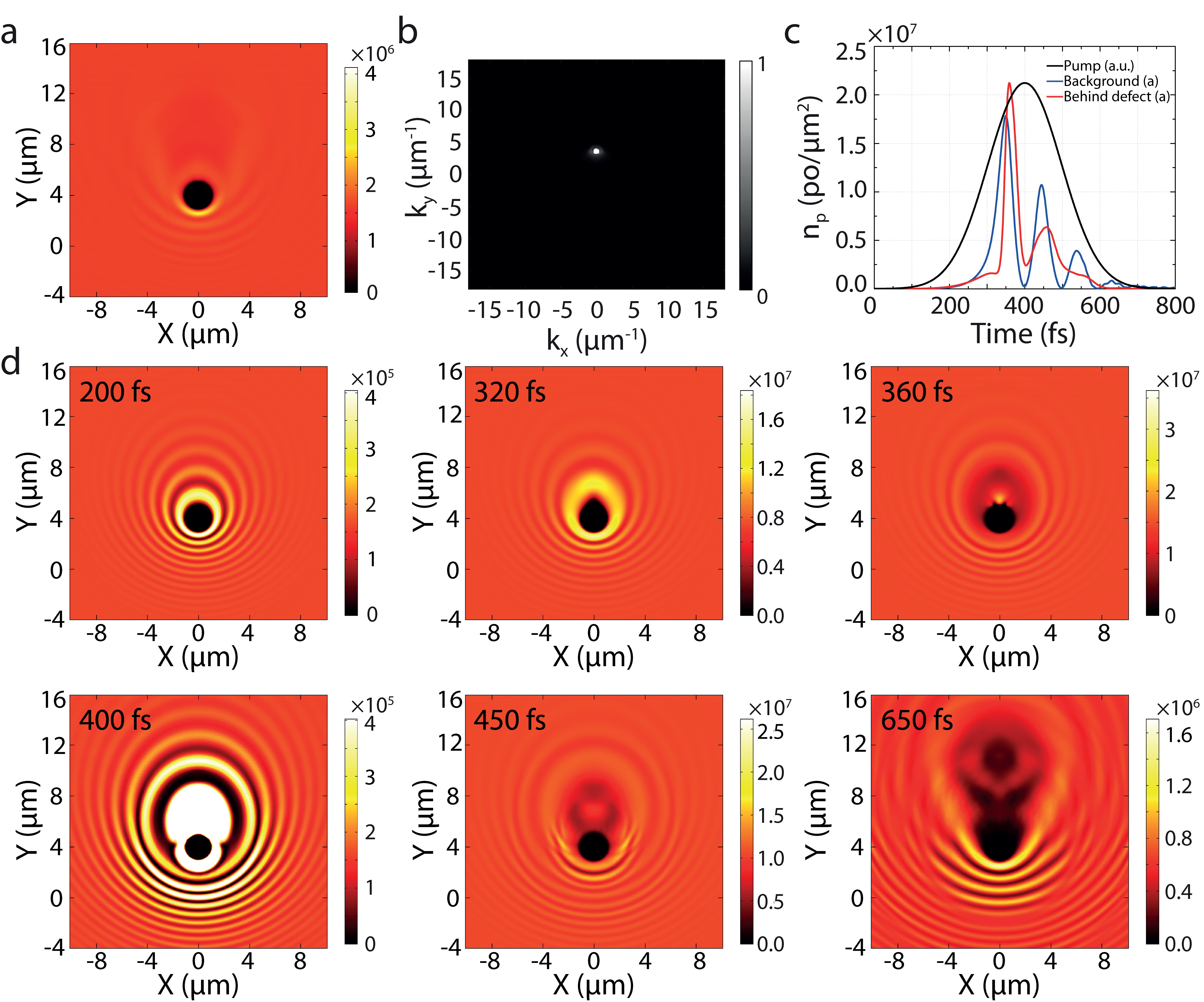} \linespread{1.1}
  \protect\protect\caption{
\textbf{Effect of pump chirp in the superfluid regime.} \textbf{a,} Real space polariton density profile in the superfluid regime when the pump is chirped by going through $3~\text{cm}$ of fused silica. Peak background density is $1.8 \cdot 10^7~\text{pol}/\upmu\text{m}^2$. \textbf{b,} Saturated intensity momentum space emission profile showing the absence of the Rayleigh scattering ring pattern. \textbf{c,} Time-domain polariton density traces behind the defect (red) and in the background (blue). The pump pulse is traced in black and normalized to the peak density. \textbf{d,} Individual time snapshots of the polariton density during polariton flow. The color scales are in $\text{pol}/\upmu\text{m}^2$ and their maximum values are adjusted to maintain the background color throughout the images. 
 }
\label{fig:FIGS9} 
\end{figure}

\end{document}